\documentstyle[12pt]{article} 
\begin{document} 
\begin{titlepage} 
\begin{flushright} 
ULB--TH 99/01\\
April 1999
\end{flushright}
\vspace*{1.6cm} 
 
\begin{center}
{\Large\bf Phenomenological evidence for the energy dependence of the
$\eta$-$\eta\prime$ mixing angle}\\ 
\vspace*{0.8cm} 

R.~Escribano\footnote{Chercheur IISN.} and 
J.-M.~Fr\`ere\footnote{Directeur de recherches du FNRS.}\\ 
\vspace*{0.2cm} 
 
{\footnotesize\it Service de Physique Th\'eorique, Universit\'e Libre de 
Bruxelles, CP 225, B-1050 Bruxelles, Belgium} 
\end{center} 
\vspace*{1.0cm} 
 
\begin{abstract}
\noindent
A phenomenological analysis on various decay processes is performed using an 
energy-dependent $\eta$-$\eta\prime$ mixing angle scheme.
Special attention is given to the electromagnetic couplings between 
lowest-lying vector and pseudoscalar mesons.
The agreement between our predictions and the experimental values is 
remarkable. This analysis opens a connection to two-angle fits in the 
$\eta$-$\eta\prime$ sector.
\end{abstract} 
\end{titlepage} 
 
\section{Introduction}
\label{intro}
The $\eta$-$\eta\prime$ mixing angle has become one of the most interesting 
$SU(3)$-breaking hadronic parameters to measure since $SU(3)$ symmetry was
proposed \cite{RPP}.
Several exhaustive analyses surveying many different processes have been
performed along the years \cite{GK,BS,BFT,BES98}. The values presented for 
the mixing angle range from $-20^\circ$ to $-10^\circ$. 
In all those previous analyses the dependence with energy of the 
$\eta$-$\eta\prime$ mixing angle has been neglected. The main reason for such
simplification is that there is not yet an established theoretical framework 
where the $\eta\prime$ could be well incorporated.
Recently, a combination of Chiral Perturbation Theory ($\chi$PT) together 
with a simultaneous expansion in $1/N_c$ (in order to take into account the
axial $U_A(1)$ anomaly) has been proposed for describing the
$\eta\prime$ dynamics \cite{LEU1,TAR}. 
In this so-called Extended-$\chi$PT (E$\chi$PT), the energy dependence
of the $\eta$-$\eta\prime$ mixing angle could be traced. However, the 
$q^2$-dependence expected from loop corrections starts at orders
that cannot be actually computed due to the proliferation of unknown 
parameters \cite{TAR2}.
In the analogous case of $\pi^0$-$\eta$ mixing, $\chi$PT alone provides a 
valuable framework where the $q^2$-dependence can be followed, showing that 
the $\pi^0$-$\eta$ mixing angle is modified by an 8\% from $q^2=m^2_\pi$ to 
$q^2=m^2_\eta$ \cite{MAL}.
In principle, there is no reason for avoiding a similar behaviour for the
$\eta$-$\eta\prime$ case.

Yet in Ref.~\cite{BFT}, an agreement with experimental data, including $J/\psi$
radiative decays, was achieved using an energy-independent mixing angle, in 
contrast with $\chi$PT where the energy-independent parametrization fails in
describing those radiative decays. This difference is due to the fact that the
approach of Refs.~\cite{BFT,AF} contains {\it ab initio} the axial anomaly.
Similarly, the E$\chi$PT includes the effect of the axial anomaly through a
perturbative expansion, allowing the presence of the pseudoscalar singlet as an
additional degree of freedom, and thus the possibility of considering an
energy dependence for the $\eta$-$\eta\prime$ mixing angle.

In this letter we extend a previous analysis of various decay 
processes assuming an energy dependence of the $\eta$-$\eta\prime$
mixing angle and check how this modification improves the results of 
Ref.~\cite{BFT}. We do not use any theoretical prejudice about
the explicit $q^2$-dependence. We simply assume that the value of the 
mixing angle at $q^2=m^2_\eta$ can differ from the value at 
$q^2=m^2_{\eta\prime}$.
In Section \ref{notation}, we shortly introduce the notation used in the
analysis. In Sec.~\ref{fit}, we compute the radiative decays
$(\eta,\eta\prime)\rightarrow\gamma\gamma$ and the ratio 
$R_{J/\psi}\equiv\Gamma(J/\psi\rightarrow\eta\prime\gamma/\eta\gamma)$
in the energy-dependent scheme
in order to obtain the preferred values for the mixing angles involved in
the analysis. Sec.~\ref{VPgamma} is devoted to the consequences of our
approach in the context of the radiative decays of lowest-lying vector 
and pseudoscalar mesons, $V\rightarrow P\gamma$ and $P\rightarrow V\gamma$,
respectively. Finally, in Sec.~\ref{conclusions}, we present our 
conclusions.

\section{Notation}
\label{notation}
The $\eta$-$\eta\prime$ mixing angle in the octet-singlet basis allowing
for an energy dependence is written as
\begin{equation}
\label{defbasis}
\begin{array}{c}
|\eta\rangle=c\theta_\eta |\eta_8\rangle-s\theta_\eta |\eta_0\rangle\ ,\\[1ex]
|\eta\prime\rangle=
s\theta_{\eta\prime} |\eta_8\rangle+c\theta_{\eta\prime} |\eta_0\rangle\ ,
\end{array}
\end{equation}
where $\theta_\eta$ and $\theta_{\eta\prime}$ are defined as the values
of the mixing angle $\theta(q^2)$ at $q^2=m^2_\eta$ and
$q^2=m^2_{\eta\prime}$ respectively. Due to the assumed energy 
dependence, the orthogonality of the physical states in 
Eq.~(\ref{defbasis}) is no longer satisfied, contrary to the usual
energy-independent scheme where 
$\theta_\eta=\theta_{\eta\prime}\equiv\theta$ is assumed \cite{RPP}.

The pseudoscalar decay constants $f_P^i\ (i=8,0; P=\eta,\eta\prime)$ are
defined as
\begin{equation}
\label{decaycon}
\langle0|A_\mu^i|P(p)\rangle=i f_P^i p_\mu\ ,
\end{equation}
where $A_\mu^{8,0}$ are the octet and singlet axial-vector currents whose
divergences are
\begin{equation}
\label{divaxial}
\begin{array}{c}
\partial^\mu A_\mu^8=\frac{2}{\sqrt{6}}
(m_u\bar u i\gamma_5 u+m_d\bar d i\gamma_5 d-
2m_s\bar s i\gamma_5 s)\ ,\\[1ex]
\partial^\mu A_\mu^0=\frac{2}{\sqrt{3}}
(m_u\bar u i\gamma_5 u+m_d\bar d i\gamma_5 d+m_s\bar s i\gamma_5 s)+
\frac{1}{\sqrt{3}}\frac{3\alpha_s}{4\pi} G_{\mu\nu}^a\tilde G^{a,\mu\nu}\ ,
\end{array}
\end{equation}
where $G_{\mu\nu}^a$ is the gluonic field-strength tensor and
$\tilde G_{\mu\nu}^a
\equiv \frac{1}{2}\epsilon_{\mu\nu\alpha\beta}G^{a,\alpha\beta}$ its dual.
The divergence of the matrix elements (\ref{decaycon}) are then written as
\begin{equation}
\label{divdecaycon}
\langle0|\partial^\mu A_\mu^i|P\rangle=f_P^i m_P^2\ ,
\end{equation}
where $m_P$ is the mass of the pseudoscalar meson.

In our analysis, we assume that the pseudoscalar decay constants will follow
the same mixing pattern as the particle state mixing does\footnote{In 
Eq.~(\ref{defdecaybasis}), we assume for definiteness that only the mixing 
angle $\theta$ (and not the decay constants $f_{8,0}$) is energy-dependent. 
In fact, the fit to the experimental data only deals with the energy 
dependence of the products $f_P^i$, and not with the energy dependence of 
$\theta$ and $f_{8,0}$ separately. A more refined analysis using for instance 
off-shell processes might help to distinguish between both energy 
dependences.} 
(see Eq.~(\ref{defbasis})):
\begin{equation}
\label{defdecaybasis}
\left(
\begin{array}{cc}
f^8_\eta & f^0_\eta \\[1ex]
f^8_{\eta\prime} & f^0_{\eta\prime}
\end{array}
\right)
=
\left(
\begin{array}{cc}
f_8 c\theta_\eta & -f_0 s\theta_\eta \\[1ex]
f_8 s\theta_{\eta\prime} &  f_0 c\theta_{\eta\prime}
\end{array}
\right)\ .
\end{equation}

Neglecting the contribution of the {\it up} and {\it down} quark masses,
as in Ref.~\cite{AF}, the matrix elements of the chiral anomaly between the 
vacuum and $(\eta, \eta\prime)$ states are
\begin{equation}
\label{chiralanomaly}
\begin{array}{c}
\langle0|\frac{3\alpha_s}{4\pi} G\tilde G|\eta\rangle=
\sqrt{\frac{3}{2}}m_\eta^2 
(f_8 c\theta_\eta-\sqrt{2} f_0 s\theta_\eta)\ ,\\[1ex]
\langle0|{\frac{3\alpha_s}{4\pi}} G\tilde G|\eta\prime\rangle=
\sqrt{\frac{3}{2}}m_{\eta\prime}^2 
(f_8 s\theta_{\eta\prime}+\sqrt{2} f_0 c\theta_{\eta\prime})\ .
\end{array}
\end{equation}

\section{Experimental values for the $\theta_\eta$ and $\theta_{\eta\prime}$
mixing angles}
\label{fit}
In order to reach some predictions from our energy-dependent mixing angle
analysis we must first know the values of $\theta_\eta$ and
$\theta_{\eta\prime}$ preferred by the experimental data. We will use as 
constraints\footnote{
We choose such constrains because those decays are well understood
in terms of the electromagnetic anomaly (see e.g.~Ref.~\cite{DGH}).} 
the experimental decay widths of 
$(\eta, \eta\prime)\rightarrow\gamma\gamma$ \cite{RPP}
\begin{equation}
\label{expwidths}
\begin{array}{c}
\Gamma(\eta\rightarrow\gamma\gamma)=(0.46\pm 0.04)\ \mbox{keV}\ ,\\[1ex]
\Gamma(\eta\prime\rightarrow\gamma\gamma)=(4.27\pm 0.19)\ \mbox{keV}\ .
\end{array}
\end{equation}
Generalizing the PCAC result for the $\pi^0\rightarrow\gamma\gamma$ decay,
one assumes that the interpolating fields $\eta$ and $\eta\prime$ can be
related via PCAC with the axial-vector currents (see e.g.~Refs.~\cite{KP,CF}) 
in the following way
\begin{equation}
\label{interfields}
\begin{array}{c}
\eta(x)=\frac{1}{m_\eta^2}
\frac{f^0_{\eta\prime}\partial^\mu A_\mu^8(x)-
      f^8_{\eta\prime}\partial^\mu A_\mu^0(x)}
{f^0_{\eta\prime}f^8_\eta-f^8_{\eta\prime}f^0_\eta}\ ,\\[2ex]
\eta\prime(x)=\frac{1}{m_{\eta\prime}^2}
\frac{f^0_\eta\partial^\mu A_\mu^8(x)-
      f^8_\eta\partial^\mu A_\mu^0(x)}
{f^0_\eta f^8_{\eta\prime}-f^8_\eta f^0_{\eta\prime}}\ .
\end{array}
\end{equation}
This leads to\footnote{Note that if one assumes an energy-independent
parameterization of the mixing angle 
$(\theta_\eta=\theta_{\eta\prime}\equiv\theta)$ the standard result is
obtained (see e.g.~Ref.~\cite{DGH}).}
\begin{equation}
\label{theowidths}
\begin{array}{c}
\Gamma(\eta\rightarrow\gamma\gamma)=
\frac{\alpha^2 m_\eta^3}{96\pi^3}\left(
\frac{f^0_{\eta\prime}-2\sqrt{2}f^8_{\eta\prime}}
{f^0_{\eta\prime}f^8_\eta-f^8_{\eta\prime}f^0_\eta}\right)^2=
\frac{\alpha^2 m_\eta^3}{96\pi^3}\left(
\frac{c\theta_{\eta\prime}/f_8-2\sqrt{2}s\theta_{\eta\prime}/f_0}
{c\theta_{\eta\prime}c\theta_\eta+
 s\theta_{\eta\prime}s\theta_\eta}\right)^2\ ,\\[2ex]
\Gamma(\eta\prime\rightarrow\gamma\gamma)=
\frac{\alpha^2 m_{\eta\prime}^3}{96\pi^3}\left(
\frac{f^0_\eta-2\sqrt{2}f^8_\eta}
{f^0_\eta f^8_{\eta\prime}-f^8_\eta f^0_{\eta\prime}}\right)^2=
\frac{\alpha^2 m_{\eta\prime}^3}{96\pi^3}\left(
\frac{s\theta_\eta/f_8+2\sqrt{2}c\theta_\eta/f_0}
{c\theta_{\eta\prime}c\theta_\eta+
 s\theta_{\eta\prime}s\theta_\eta}\right)^2\ .
\end{array}
\end{equation}
Because of the four unknown parameters 
($\theta_\eta, \theta_{\eta\prime}, f_8$ and $f_0$) that appear in 
Eq.~(\ref{theowidths}), in order to get their allowed values we need two 
additional constraints (apart from the experimental constraints in 
Eq.~(\ref{expwidths})).
On the one hand, we will use the well established prediction of 
$\chi$PT: $f_8=1.28 f_\pi$ ($f_\pi$= 132 MeV) as a theoretical constrain. 
On the other hand, we will use the experimental value of the ratio \cite{RPP}
\begin{equation}
\label{RJpsiexp}
R_{J/\psi}\equiv
\frac{\Gamma(J/\psi\rightarrow\eta\prime\gamma)}
     {\Gamma(J/\psi\rightarrow\eta\gamma)}
=5.0\pm 0.6\ .
\end{equation}
According to Ref.~\cite{NOV}, the radiative $J/\psi\rightarrow P\gamma$ decays
are dominated by non-perturbative gluonic matrix elements (see Ref.~\cite{BFT}
for further comments on the accuracy of this statement):
\begin{equation}
\label{RJpsitheo}
R_{J/\psi}=\left|
\frac{\langle 0|G\tilde G|\eta\prime\rangle}{\langle 0|G\tilde G|\eta\rangle}
\right|^2\left(\frac{p_{\eta\prime}}{p_\eta}\right)^3\ ,
\end{equation}
where $p_P=M_{J/\psi}(1-m_P^2/M_{J/\psi}^2)/2$ is the three-momentum of the 
$P$-meson in the rest frame of the decaying $J/\psi$ (with mass $M_{J/\psi}$). 
Using Eq.~(\ref{chiralanomaly}) one gets
\begin{equation}
\label{RJpsiourtheo}
R_{J/\psi}=\left|
\frac{m_{\eta\prime}^2(f^8_{\eta\prime}+\sqrt{2}f^0_{\eta\prime})}
     {m_\eta^2(f^8_\eta+\sqrt{2}f^0_\eta)}
\right|^2\left(\frac{p_{\eta\prime}}{p_\eta}\right)^3=\left(
\frac{
m_{\eta\prime}^2(f_8 s\theta_{\eta\prime}+\sqrt{2}f_0 c\theta_{\eta\prime})}
{m_\eta^2(f_8 c\theta_\eta-\sqrt{2}f_0 s\theta_\eta)}
\right)^2\left(\frac{p_{\eta\prime}}{p_\eta}\right)^3\ .
\end{equation}
Comparing the experimental values of 
$\Gamma(\eta,\eta\prime\rightarrow\gamma\gamma)$ and $R_{J/\psi}$ with the
theoretical predictions shown in Eqs.~(\ref{theowidths}) and 
(\ref{RJpsiourtheo}), one obtains
\begin{equation}
\label{fitresult}
\begin{array}{c}
\theta_\eta=(-6.5\pm 2.5)^\circ\ ,\\[1ex]
\theta_{\eta\prime}=(-23.1\pm 3.0)^\circ\ ,\\[1ex] 
f_0=(1.31\pm 0.07) f_\pi\ .
\end{array}
\end{equation}
The previous values constitute the first result of the present analysis.
It is worth noting that the $\theta_\eta$ and $\theta_{\eta\prime}$ 
mixing angle values are different at the $3\sigma$ level, while the value of
the pseudoscalar decay constant $f_0$ remains compatible with other results.
Eq.~(\ref{fitresult}) shows that the difference between the value of the 
mixing angle at $q^2=m^2_\eta$ and $q^2=m^2_{\eta\prime}$ is of the order of 
250\%, a huge difference compared to the 8\% corresponding to the 
$\pi^0$-$\eta$ $q^2$-dependence. 
We are not able to argue the reason for such a different
behaviour because the values presented in Eq.~(\ref{fitresult}) are 
the result of a phenomenological analysis. 
Nevertheless, we must make some comments that could be relevant for 
understanding the origin of that difference:
on the one hand, we assume that all the energy dependence is included in the
mixing angle (see Sec.~\ref{notation}).
On the other hand, the $\eta$-$\eta\prime$ sector is different from the 
$\pi^0$-$\eta$ sector because of the major r\^ole played by the axial anomaly
in the singlet mass (from which the mixing angle is defined), that prevents
the singlet from being a Goldstone boson in the chiral limit. This fact
induces a different behaviour on the $\eta$-$\eta\prime$ mixing angle, and
presumably on its energy dependence, with respect to its $\pi^0$-$\eta$
counterpart. However, the definite resolution of this dilemma would come from
the analytical calculation of the energy dependence of the mixing angle in the
framework of the E$\chi$PT, that would provide a valuable comparison with 
phenomenological analyses, comparison that is already available for the
$\pi^0$-$\eta$ case.

Recently, in a different approach based on the E$\chi$PT, it was 
proven \cite{LEU1,LEU2} that at next-to-leading order the 
pseudoscalar decay constants $f_P^i$ could not be explained in
terms of just one mixing angle, but two different mixing angles would be  
needed, one associated to the octet and the other to the 
singlet\footnote{The original notation in Refs.~\cite{LEU1,LEU2} for the
octet and singlet decay constants $f_8$ and $f_0$ is slightly modified here
to $\tilde f_8$ and $\tilde f_0$ in order to avoid a confusion
between the two schemes (see Eqs.~(\ref{defdecaybasis}) and 
(\ref{defdecaybasisECHPT})).}:
\begin{equation}
\label{defdecaybasisECHPT}
\left(
\begin{array}{cc}
f^8_\eta & f^0_\eta \\[1ex]
f^8_{\eta\prime} & f^0_{\eta\prime}
\end{array}
\right)
=
\left(
\begin{array}{cc}
\tilde f_8 c\theta_8 & -\tilde f_0 s\theta_0 \\[1ex]
\tilde f_8 s\theta_8 &  \tilde f_0 c\theta_0
\end{array}
\right)\ .
\end{equation}
In that theoretical framework one gets ($\tilde f_8=1.28 f_\pi$ fixed from 
$\chi$PT)
\begin{equation}
\label{ECHPTresult}
\theta_8=-20.5^\circ\ ,\hspace{0.6cm} 
\theta_0\simeq -4^\circ\ ,\hspace{0.6cm}
\tilde f_0\simeq 1.25 f_\pi\ .
\end{equation}
Ref.~\cite{FK} uses the hypothesis that the four wave functions associated to 
the pseudoscalar mesons, from which the pseudoscalar decay constants are 
defined, are different. Then, two different mixing angles for the decay 
constants must be introduced. Their analysis gives
\begin{equation}
\label{FKresult}
\theta_8=-22.2^\circ\ ,\hspace{0.6cm} 
\theta_0\simeq -9.1^\circ\ ,\hspace{0.6cm}
\tilde f_0\simeq 1.20 f_\pi\ .
\end{equation}
It is straightforward to establish the link between our results in 
Eq.~(\ref{fitresult}) and the octet-singlet two-angle mixing scheme for the
pseudoscalar decay constants
\begin{equation}
\label{changeangles}
\begin{array}{c}
\tan\theta_8=\frac{\sin\theta_{\eta\prime}}{\cos\theta_\eta}\ ,\hspace{1.2cm}
\tan\theta_0=\frac{\sin\theta_\eta}{\cos\theta_{\eta\prime}}\ ,\\[2ex]
\tilde f_0=f_0\sqrt{s^2\theta_\eta+c^2\theta_{\eta\prime}}\ ,
\end{array}
\end{equation}
yielding
\begin{equation}
\label{fitresult80}
\theta_8=(-21.5\pm 2.4)^\circ\ ,\ 
\theta_0=(-7.0\pm 2.7)^\circ\ ,\ 
\tilde f_0=(1.21\pm 0.07) f_\pi\ .
\end{equation}
These values are fully compatible with those obtained in Refs.~\cite{LEU1,FK}.
The pseudoscalar decay constants $f^8_{\eta,\eta\prime}$ and 
$f^0_{\eta,\eta\prime}$ will be the same in both approaches since they are 
directly related to the physical measurements, while the inferred 
(theoretical) quantities $\tilde f_{8,0}$ and $f_{8,0}$ may differ in the 
fits due to the different approach taken.

As far as two-angle fits are concerned, we must stress here that the two
mixing angles $\theta_8$ and $\theta_0$ introduced in Refs.~\cite{LEU1,LEU2,FK}
to parameterize the pseudoscalar decay constants are conceptually 
very different to the particle state mixing angle $\theta$ that rotates the
flavour states $(\eta_8,\eta_0)$ into the physical states $(\eta,\eta\prime)$.
However, because in our energy-dependent mixing angle scheme we do not 
distinguish between the mixing properties of the meson states from the 
mixing properties of the decay constants (see Sec.~\ref{notation}), a 
description of $\theta_8$ and $\theta_0$ in terms of $\theta_\eta$ and 
$\theta_{\eta\prime}$, as done in Eq.~(\ref{changeangles}), 
is allowed\footnote{The same statement in the usual
energy-independent mixing angle scheme yields $\theta_8=\theta_0=\theta$ and
$\tilde f_{8,0}=f_{8,0}$.}.

\section{$V$-$P$ electromagnetic form factors in the energy-dependent mixing 
angle scheme}
\label{VPgamma}
In this section, we want to test some of the consequences of our approach.
In particular, we are interested in the couplings of the radiative decays of
lowest-lying vector mesons, $V\rightarrow(\eta,\eta\prime)\gamma$, and of the 
radiative decays $\eta\prime\rightarrow V\gamma$, with $V=\rho, \omega, \phi$.
In order to predict such couplings we follow closely the method 
presented in Ref.~\cite{BFT} where the description of the light vector 
meson decays is based on their relation with the $AVV$ triangle anomaly, $A$ 
and $V$ being an axial-vector and a vector current respectively.
The approach both includes $SU_F(3)$ breaking effects and fixes the vertex
couplings $g_{VP\gamma}$ as explained below.

In that framework, one starts considering the correlation function
\begin{equation}
\label{corrfun}
i\int d^4x e^{iq_1 x}
\langle P(q_1+q_2)|TJ_\mu^{\rm EM}(x)J_\nu^V(0)|0\rangle=
\epsilon_{\mu\nu\alpha\beta}q_1^\alpha q_2^\beta F_{VP\gamma}(q_1^2,q_2^2)\ ,
\end{equation}
where the currents are defined as
\begin{equation}
\label{defcurr}
\begin{array}{c}
J_\mu^{\rm EM}=\frac{2}{3}\bar u\gamma_\mu u-\frac{1}{3}\bar d\gamma_\mu d-
               \frac{1}{3}\bar s\gamma_\mu s\ ,\\[1ex]
J_\mu^{\rho,\omega}=
\frac{1}{\sqrt{2}}(\bar u\gamma_\mu u\pm\bar d\gamma_\mu d)\hspace{0.6cm}
\mbox{and}\hspace{0.6cm} J_\mu^\phi=-\bar s\gamma_\mu s\ .
\end{array}
\end{equation}
The form factors values $F_{VP\gamma}(0,0)$ are fixed by the $AVV$ triangle 
anomaly (one $V$ being an electromagnetic current), and are written in terms 
of the pseudoscalar decay constants and the $\phi$-$\omega$ mixing angle 
$\theta_V$ as
\begin{equation}
\label{FVPgamma00}
\begin{array}{l}
F_{\rho\eta\gamma}(0,0)=\frac{\sqrt{3}}{4\pi^2}
\frac{f^0_{\eta\prime}-\sqrt{2}f^8_{\eta\prime}}
{f^0_{\eta\prime}f^8_\eta-f^8_{\eta\prime}f^0_\eta}\ ,\\[1ex]
F_{\rho\eta\prime\gamma}(0,0)=\frac{\sqrt{3}}{4\pi^2}
\frac{f^0_\eta-\sqrt{2}f^8_\eta}
{f^0_\eta f^8_{\eta\prime}-f^8_\eta f^0_{\eta\prime}}\ ,\\[1ex]
F_{\omega\eta\gamma}(0,0)=\frac{1}{2\sqrt{2}\pi^2}
\frac{(c\theta_V-s\theta_V/\sqrt{2})f^0_{\eta\prime}-
      s\theta_V f^8_{\eta\prime}}
{f^0_{\eta\prime}f^8_\eta-f^8_{\eta\prime}f^0_\eta}\ ,\\[1ex]
F_{\omega\eta\prime\gamma}(0,0)=\frac{1}{2\sqrt{2}\pi^2}
\frac{(c\theta_V-s\theta_V/\sqrt{2})f^0_\eta-s\theta_V f^8_\eta}
{f^0_\eta f^8_{\eta\prime}-f^8_\eta f^0_{\eta\prime}}\ ,\\[1ex]
F_{\phi\eta\gamma}(0,0)=-\frac{1}{2\sqrt{2}\pi^2}
\frac{(s\theta_V+c\theta_V/\sqrt{2})f^0_{\eta\prime}+
      c\theta_V f^8_{\eta\prime}}
{f^0_{\eta\prime}f^8_\eta-f^8_{\eta\prime}f^0_\eta}\ ,\\[1ex]
F_{\phi\eta\prime\gamma}(0,0)=-\frac{1}{2\sqrt{2}\pi^2}
\frac{(s\theta_V+c\theta_V/\sqrt{2})f^0_\eta+c\theta_V f^8_\eta}
{f^0_\eta f^8_{\eta\prime}-f^8_\eta f^0_{\eta\prime}}\ .\\[1ex]
\end{array}
\end{equation}
Using their analytic properties, we can express these form factors by a 
dispersion relation in the momentum of the vector current, which are then
saturated with the lowest-lying resonances:
\begin{equation}
\label{VMD}
F_{VP\gamma}(0,0)=\frac{f_V}{m_V}g_{VP\gamma}+\cdots\ ,
\end{equation}
where the dots stand for higher resonances and multiparticle contributions
to the correlation function. In the following we assume vector meson 
dominance (VMD) and thus neglect these contributions (see Ref.~\cite{BFT} 
for further details).

The $f_V$ are the vector mesons' leptonic decay constants defined by
\begin{equation}
\label{lepdeccon}
\langle0|J_\mu^V|V(p,\lambda)\rangle=m_V f_V \varepsilon_\mu^{(\lambda)}(p)\ ,
\end{equation}
where $m_V$ and $\lambda$ are the mass and the helicity state of the vector 
meson. The $f_V$ can be determined from the experimental decay 
rates \cite{RPP} via
\begin{equation}
\label{GVee}
\Gamma(V\rightarrow e^+e^-)=\frac{4\pi}{3}\alpha^2\frac{f_V^2}{m_V}c_V^2\ ,
\end{equation}
with $c_V
=(\frac{1}{\sqrt{2}},\frac{s\theta_V}{\sqrt{6}},\frac{c\theta_V}{\sqrt{6}})$
for $V=\rho, \omega, \phi$.  The experimental values are
\begin{equation}
\label{fVexp}
\begin{array}{c}
f_{\rho^0}=(216\pm 5)\ \mbox{MeV}\ ,\\[1ex]
f_{\omega}=(180\pm 3)\ \mbox{MeV}\ ,\\[1ex]
f_{\phi}=(244\pm 4)\ \mbox{MeV}\ .
\end{array}
\end{equation}

Finally, we introduce the vertex couplings $g_{VP\gamma}$, which are just the
on-shell $V$-$P$ electromagnetic form factors:
\begin{equation}
\label{gVPgamma}
\langle P(p_P)|J_\mu^{\rm EM}|V(p_V,\lambda)\rangle|_{(p_V-p_P)^2=0}=
-g_{VP\gamma}\epsilon_{\mu\nu\alpha\beta}
p_P^\nu p_V^\alpha\varepsilon_V^\beta(\lambda)\ .
\end{equation}
The decay widths of $P\rightarrow V\gamma$ and $V\rightarrow P\gamma$ are
\begin{equation}
\label{GVPgamma}
\begin{array}{l}
\Gamma(P\rightarrow V\gamma)=
\frac{\alpha}{8}g_{VP\gamma}^2\left(\frac{m_P^2-m_V^2}{m_P}\right)^3\ ,\\[1ex]
\Gamma(V\rightarrow P\gamma)=
\frac{\alpha}{24}g_{VP\gamma}^2\left(\frac{m_V^2-m_P^2}{m_V}\right)^3\ .
\end{array}
\end{equation}

\begin{table}
\centering
{\scriptsize
\begin{tabular}{lllll}
\hline\hline\\
$V$ & $P$ & $g_{VP\gamma}$ (th.) & & $g_{VP\gamma}$ (exp.) \\[2ex]
\hline\\
$\rho$ & $\eta$ & 
$\frac{\sqrt{3} m_\rho}{4\pi^2 f_\rho}
\frac{c\theta_{\eta\prime}/f_8-\sqrt{2}s\theta_{\eta\prime}/f_0}
{c\theta_{\eta\prime}c\theta_\eta+s\theta_{\eta\prime}s\theta_\eta}$
& $=(1.43\pm 0.10)$ GeV$^{-1}$ & $(1.47\pm 0.28)$ GeV$^{-1}$ \\[1ex]
$\rho$ & $\eta\prime$ &
$\frac{\sqrt{3} m_\rho}{4\pi^2 f_\rho}
\frac{s\theta_\eta/f_8+\sqrt{2}c\theta_\eta/f_0}
{c\theta_{\eta\prime}c\theta_\eta+s\theta_{\eta\prime}s\theta_\eta}$
& $=(1.23\pm 0.11)$ GeV$^{-1}$ & $(1.31\pm 0.06)$ GeV$^{-1}$ \\[1ex]
$\omega$ & $\eta$ &
$\frac{m_\omega}{2\sqrt{2}\pi^2 f_\omega}
\frac{(c\theta_V-s\theta_V/\sqrt{2})c\theta_{\eta\prime}/f_8-
      s\theta_V s\theta_{\eta\prime}/f_0}
{c\theta_{\eta\prime}c\theta_\eta+s\theta_{\eta\prime}s\theta_\eta}$
& $=(0.54\pm 0.04)$ GeV$^{-1}$ & $(0.53\pm 0.04)$ GeV$^{-1}$ \\[1ex]
$\omega$ & $\eta\prime$ & 
$\frac{m_\omega}{2\sqrt{2}\pi^2 f_\omega}
\frac{(c\theta_V-s\theta_V/\sqrt{2})s\theta_\eta/f_8+
      s\theta_V c\theta_\eta/f_0}
{c\theta_{\eta\prime}c\theta_\eta+s\theta_{\eta\prime}s\theta_\eta}$
& $=(0.55\pm 0.05)$ GeV$^{-1}$ & $(0.45\pm 0.03)$ GeV$^{-1}$ \\[1ex]
$\phi$ & $\eta$ & 
$-\frac{m_\phi}{2\sqrt{2}\pi^2 f_\phi}
\frac{(s\theta_V+c\theta_V/\sqrt{2})c\theta_{\eta\prime}/f_8+
      c\theta_V s\theta_{\eta\prime}/f_0}
{c\theta_{\eta\prime}c\theta_\eta+s\theta_{\eta\prime}s\theta_\eta}$ 
& $=(0.73\pm 0.06)$ GeV$^{-1}$ & $(0.69\pm 0.02)$ GeV$^{-1}$ \\[1ex]
$\phi$ & $\eta\prime$ & 
$-\frac{m_\phi}{2\sqrt{2}\pi^2 f_\phi}
\frac{(s\theta_V+c\theta_V/\sqrt{2})s\theta_\eta/f_8-
      c\theta_V c\theta_\eta/f_0}
{c\theta_{\eta\prime}c\theta_\eta+s\theta_{\eta\prime}s\theta_\eta}$
& $=(0.83\pm 0.06)$ GeV$^{-1}$ & $(1.01\pm 0.30)$ GeV$^{-1}$ \\[2ex]
\hline\hline
\end{tabular}
}
\caption{Theoretical and experimental values of the on-shell 
$V$-$(\eta,\eta\prime)$ electromagnetic vertex couplings in the 
energy-dependent $\eta$-$\eta\prime$ mixing angle scheme. 
For $g_{VP\gamma}$ (th.) we give the experimental errors coming from the 
decay constants $f_{P,V}$ and the mixing angle values $\theta_\eta$ and
$\theta_{\eta\prime}$. We use $\theta_V=38.6^\circ$ for the $\phi$-$\omega$ 
mixing angle. Experimental values are taken from {\protect\cite{RPP}}.}
\label{table1}
\end{table}

Eq.~(\ref{VMD}) allows us to identify the $g_{VP\gamma}$ couplings defined in
(\ref{gVPgamma}) with the form factors $F_{VP\gamma}(0,0)$ listed in
(\ref{FVPgamma00}). We make use of this relationship to predict these
couplings in the energy-dependent $\eta$-$\eta\prime$ mixing angle scheme.
Our theoretical expectations are shown in Table \ref{table1}.
The couplings are expressed in terms of the mixing 
angle values $\theta_\eta$ and $\theta_{\eta\prime}$, the octet and singlet
pseudoscalar decay constants $f_8$ and $f_0$, the $\phi$-$\omega$ mixing 
angle $\theta_V$, and the corresponding vector decay constants $f_V$. We also
include a numerical prediction for each coupling that should be compared with
the experimental values extracted from (\ref{GVPgamma}) and \cite{RPP}.
In the numerical analysis we have included the deviation from ideal
$\phi$-$\omega$ mixing by taking into account a value for the mixing angle of
$\theta_V=38.6^\circ$ \cite{RPP}. 
The error quoted in Table \ref{table1} does not reflect the full theoretical 
uncertainty, but namely propagates the errors from (\ref{fitresult}) and
(\ref{fVexp}). The agreement between our predictions and
the experimental values is quite remarkable: all the values coincide at the
$1\sigma$ level except for the $g_{\omega\eta\prime\gamma}$ case. However,
the $g_{\omega\eta\prime\gamma}$ coupling is rather sensitive to the
$\phi$-$\omega$ mixing angle; for instance setting $\theta_V$ to the ideal
mixing value of $35.3^\circ$ reduces the coupling by a 9\%. As seen from 
the table, the prediction for the decay $\phi\rightarrow\eta\gamma$ 
(which is largely independent of $\theta_V$) is in clear agreement with data, 
contrary to the encountered situation in the energy-independent mixing scheme 
where the $g_{\phi\eta\gamma}$ is only consistent with the experimental value 
for $\theta\simeq -22^\circ$ \cite{BFT}. The improvement over Ref.~\cite{BFT}
is obvious in this case. We also give a precise prediction for 
the $\phi\rightarrow\eta\prime\gamma$ decay, which is compatible with current 
data, although the experimental error is still too big to be conclusive.

\begin{table}
\centering
\begin{tabular}{cccc}
\hline\hline\\
Assumptions & Fitted data & Results & $\chi^2/dof$ \\[2ex]
\hline\hline\\
$\theta_\eta=\theta_{\eta\prime}\equiv \theta$ &
All data & $\theta=(-18.1\pm 1.2)^\circ$ & 22.2/7 \\[1ex]
$f_8=1.28 f_\pi$ & & $f_0=(1.13\pm 0.03) f_\pi$ & \\[1ex]\hline\\[-1ex]
$\theta_\eta=\theta_{\eta\prime}\equiv \theta$ &
All data & $\theta=(-18.1\pm 1.2)^\circ$ & 22.2/6 \\[1ex]
$f_8$ free & & $f_8=(1.28\pm 0.04) f_\pi$ & \\[1ex]
 & & $f_0=(1.13\pm 0.03) f_\pi$ & \\[1ex]\hline\hline\\[-1ex]
$\theta_\eta\neq \theta_{\eta\prime}$ &
$\Gamma(\eta,\eta\prime\rightarrow\gamma\gamma)$ + &
$\theta_\eta=(-6.5\pm 2.5)^\circ$ & not applicable \\[1ex]
$f_8=1.28 f_\pi$ & $R_{J/\psi}$ &
$\theta_{\eta\prime}=(-23.1\pm 3.0)^\circ$ & \\[1ex]
 & & $f_0=(1.31\pm 0.07) f_\pi$ & \\[1ex]\hline\\[-1ex]
$\theta_\eta\neq \theta_{\eta\prime}$ &
All data & $\theta_\eta=(-6.9\pm 2.1)^\circ$ & 8.4/6 \\[1ex]
$f_8=1.28 f_\pi$ & & $\theta_{\eta\prime}=(-24.6\pm 2.3)^\circ$ & \\[1ex]
 & & $f_0=(1.34\pm 0.07) f_\pi$ & \\[1ex]\hline\\[-1ex]
$\theta_\eta\neq \theta_{\eta\prime}$ &
All data & $\theta_\eta=(-5.7\pm 2.7)^\circ$ & 7.9/5 \\[1ex]
$f_8$ free & & $\theta_{\eta\prime}=(-24.6\pm 2.3)^\circ$ & \\[1ex]
 & & $f_8=(1.32\pm 0.06) f_\pi$ & \\[1ex]
 & & $f_0=(1.37\pm 0.07) f_\pi$ & \\[2ex]
\hline\hline
\end{tabular}
\caption{Results for the $\eta$-$\eta\prime$ mixing angle and decay constants
in the energy-independent and -dependent mixing angle schemes. For every fit,
the theoretical assumptions taken, the set of fitted experimental data, and
the value of the $\chi^2/dof$ are shown in the first, second and last column
respectively. Numerical results are presented in the third column. All data
stands for the decay widths of $(\eta,\eta\prime)\rightarrow\gamma\gamma$, 
$V\rightarrow P\gamma$, $P\rightarrow V\gamma$, and the ratio $R_{J/\psi}$.}
\label{table2}
\end{table}

Table \ref{table1} constitutes the main result of our work.
Our analysis shows that the assumption of saturating the form factors
$F_{VP\gamma}$ by lowest-lying resonances is satisfactory (a conclusion 
already reached in Ref.~\cite{BFT}), and that the energy-dependent
$\eta$-$\eta\prime$ mixing angle scheme fits the data somewhat better than the 
energy-independent scheme does. To quantify this improvement, we have performed
various fits to the full set of experimental data assuming, or not, the energy 
dependence of the $\eta$-$\eta\prime$ mixing angle. 
The results are presented in Table \ref{table2}.
To check the consistency of our approach we have extended the fit in 
Eq.~(\ref{fitresult}) (third data row in Table \ref{table2}) to include
all experimental data that account not only for the decay widths 
$(\eta,\eta\prime)\rightarrow\gamma\gamma$ and the ratio $R_{J/\psi}$ but also
the radiative decay widths of $V\rightarrow P\gamma$ and
$P\rightarrow V\gamma$. The theoretical constraint $f_8=1.28 f_\pi$ is also
relaxed in order to test the stability of the result.
As seen from Table \ref{table2}, a significant enhancement in 
the $\chi^2/dof$ is achieved when the constrain 
$\theta_\eta=\theta_{\eta\prime}$ is relaxed, allowing us to show explicitly
the improvement of our analysis with respect to the one in Ref.~\cite{BFT}.

\section{Conclusions}
\label{conclusions}
In this letter, we have performed a phenomenological analysis on various 
decay processes assuming an energy dependence of the $\eta$-$\eta\prime$
mixing angle as the main work hypothesis.
We have been mainly interested in the consequences of such hypothesis on the 
electromagnetic couplings of $V\rightarrow P\gamma$ and 
$P\rightarrow V\gamma$ processes.
The radiative decays $(\eta,\eta\prime)\rightarrow\gamma\gamma$ together with
the ratio $R_{J/\psi}$ have been used to fit the values of the mixing angle
and pseudoscalar decay constants. Using the description of vector meson decays
in terms of their relation with the $AVV$ triangle anomaly, a theoretical
prediction for the $g_{VP\gamma}$ couplings have been derived. The agreement
between our theoretical predictions and the experimental values is quite 
remarkable and can be considered as a consistency check of the whole approach.
We have also established the link with the two-angle approaches of 
Refs.~\cite{LEU1,LEU2,FK}.

\section{Acknowledgments} 
This work was supported by the IISN (Belgium) and by the Communaut\'e
Fran\c caise de Belgique (Direction de la Recherche Scientifique 
programme ARC).


\begin{thebibliography}{99}
\bibitem{RPP}
{\it Review of Particle Properties}, Particle Data Group, 
C. Caso {\it et al.}, Eur. Phys. J. {\bf C3}, 1 (1998).
\bibitem{GK}
F. J. Gilman and R. Kauffman, Phys. Rev. {\bf D36}, 2761 (1987).
\bibitem{BS}
A. Bramon and M. D. Scadron, Phys. Lett. {\bf B234}, 346 (1990).
\bibitem{BFT}
P. Ball, J.-M. Fr\`ere and M. Tytgat, Phys. Lett. {\bf B365}, 367 (1996).
\bibitem{BES98}
A. Bramon, R. Escribano and M. D. Scadron, Phys. Lett. {\bf B403}, 339 (1997),
and Eur. Phys. J. {\bf C7}, 271 (1999).
\bibitem{LEU1}
H. Leutwyler, Nucl. Phys. Proc. Suppl. {\bf 64}, 223 (1998).
\bibitem{TAR}
P. Herrera-Sikl\'ody {\it et al.}, Nucl. Phys. {\bf B497}, 345 (1997).
\bibitem{TAR2}
P. Herrera-Sikl\'ody {\it et al.}, Phys. Lett. {\bf B419}, 326 (1998).
\bibitem{MAL} 
K. Maltman, Phys. Lett. {\bf B313}, 203 (1993).
\bibitem{AF}
R. Akhoury and J.-M. Fr\`ere, Phys. Lett. {\bf B220}, 258 (1989).
\bibitem{DGH}
J. F. Donoghue, E. Golowich and B. R. Holstein, 
{\it Dynamics of the Standard Model}, Cambridge University Press (1992).
\bibitem{KP}
A. V. Kisselev and V. A. Petrov, Z. Phys. {\bf C58}, 595 (1993).
\bibitem{CF}
P. Castoldi and J.-M.~Fr\`ere, Z. Phys. {\bf C40}, 283 (1988).
\bibitem{NOV}
V. A. Novikov, M. A. Shifman, A. I. Vainshtein and V. I. Zakharov,
Nucl. Phys. {\bf B165}, 55 (1980).
\bibitem{LEU2}
R. Kaiser and H. Leutwyler, hep-ph/9806336.
\bibitem{FK}
Th. Feldmann and P. Kroll, Eur. Phys. J. {\bf C5}, 327 (1998).
\end{thebibliography}
\end{document}